\documentclass{emulateapj}

\usepackage{epstopdf}
\usepackage{appendix}

\begin{document}

\title{Coronal Mass Ejection Mass, Energy, and Force Estimates using \emph{STEREO}}

\author{Eoin P. Carley \altaffilmark{1}, R. T. James McAteer  \altaffilmark{2}, and Peter  T. Gallagher  \altaffilmark{1}}
\altaffiltext{1}{Astrophysics Research Group, School of Physics, Trinity College Dublin, Dublin 2, Ireland.}
\altaffiltext{2}{ Department of Astronomy, New Mexico State University, Las Cruces, New Mexico 88003-8001, USA.}


\begin{abstract}
Understanding coronal mass ejection (CME) energetics and dynamics has been a long-standing problem, and although previous observational 
estimates have been made, such studies have been hindered by large uncertainties in CME mass.
Here, the two vantage points of the \emph{Solar Terrestrial Relations Observatory (STEREO)} COR1 and COR2 coronagraphs were used to  
accurately estimate the mass of the 2008 December 12 CME. Acceleration estimates derived from the position of the CME front in 3-D were 
combined with the mass estimates to calculate the magnitude of the kinetic energy and driving force at different stages of the CME evolution.
The CME asymptotically approaches a mass of $3.4\,\pm\,1.0\times10^{15}$\,g beyond $\sim$10\,$R_{\odot}$. 
The kinetic energy shows an initial rise towards $6.3\,\pm\,3.7\times10^{29}$\,erg at $\sim$3\,$R_{\odot}$, beyond which it rises steadily to 
$4.2\,\pm\,2.5\times10^{30}$\,erg at $\sim$18\,$R_{\odot}$.
The dynamics are described by an early phase of strong acceleration, dominated by a force of peak magnitude of $3.4\,\pm\,2.2\times10^{14}$\,N 
at $\sim$3\,$R_{\odot}$, after which a force of $3.8\,\pm\,5.4\times10^{13}$\,N takes affect between $\sim$7--18\,$R_{\odot}$. These results are 
consistent with magnetic (Lorentz) forces acting at heliocentric distances of $\lesssim$7\,$R_{\odot}$, while solar wind drag forces dominate at 
larger distances ($\gtrsim$7\,$R_{\odot}$).
\end{abstract}



\section{Introduction}
Despite many years of study, the origin of the forces that drive coronal mass ejections (CMEs) in the solar corona and interplanetary space are not 
well understood. 
From an observational viewpoint a complete understanding of CME kinematics, dynamics and forces requires not only a study of CME speed, 
acceleration and expansion but also an accurate knowledge of CME mass.  The measurements of CME mass combined with acceleration 
measurements can be used to quantify the magnitude of the force that drives a CME. Knowledge of this force magnitude can lead to an 
identification of the possible origin of the CME driver. 

There are numerous theoretical models that attempt to explain the triggering of CME eruption and its consequent propagation. Each describe the 
destabilization and propagation of a complex magnetic structure, such as a flux rope, via mechanisms that include the catastrophe model \citep
{forbes1991,forbes1995,lin2000}, magnetic breakout model \citep{antio99,lynch2008}, or a toroidal instability model \citep{chen1996,kleim2006}. 
The loss of equilibrium induced by such mechanisms results in CME propagation into interplanetary space. The predictions of these models have 
been investigated in observational studies whereby the CME kinematics are used to constrain what forces might be at play and hence which model 
best describes CME propagation. Such studies show that early phase propagation can be reasonably described by the existing models (or a 
combination of them) involving some form of magnetic CME driver \citep{manoh2003, chen2006, Schrij2008, lin2010}, and that aerodynamic drag 
of the solar wind may have a significant role at later stages of CME propagation \citep{howard2007, malo10, byr10}. Comparisons between 
modeling and observational estimates of the forces that drive CMEs requires an accurate determination of CME kinematics properties as well as 
CME mass.

To date, the most prevalent method of determining CME mass has been through the use of white light coronagraph imagers, such as the Large 
Angle Spectroscopic Coronagraph  \citep[LASCO;][]{bru95} on board the \emph{Solar and Heliospheric Observatory} \citep[\emph{SOHO};][]
{dom95}  and the twin Sun Earth Connection Coronal and Heliospheric Investigation (SECCHI) COR1 and COR2 coronagraphs \citep{how08} on 
board the \emph{Solar Terrestrial Relations Observatory}  \citep[\emph{STEREO};][]{kai08}. The white-light emission imaged by such coronagraphs 
occurs via Thomson scattering of photospheric light by coronal electrons \citep{min30, vdeh50, bil66}, the so called K-corona.  From classical 
Thomson scattering theory, the intensity of the light detected by an observer depends on the particle density of the scattering plasma. Hence, any 
density enhancement, such as a CME, over the background coronal density appears as enhanced emission in white light. The enhanced emission 
allows for a calculation of the total electron content and hence mass. 

Some of the first measurement of CME mass using scattering theory were carried out by \citet{munro1979} and \citet{poland1981} using space-based 
white light coronagraphs on board \emph{SkyLab} and U.S. military satellite\,\emph{P78-1}.  Both the early studies  and later statistical 
investigations determined that the majority of CMEs have masses in the range of 10$^{13}$--10$^{16}$\,g, \citep{vourlidas02, vour2010}. However, 
due to only a single viewpoint of observation, the longitudinal angle at which the CME propagates outwards was largely unknown in these studies 
and it is generally assumed that the CME propagates perpendicular to the observers line-of-sight (LOS). There is also the added assumption that all 
CME mass lies in the two-dimensional plane-of-sky (POS). Such assumptions can lead to a mass underestimation of up to 50\% or more \citep
{vou00}. More recent studies have employed the two viewpoint capabilities of the \emph{STEREO} mission to determine the mass of numerous 
CMEs with much less uncertainty \citep{cola09}. 
	
In this paper, we analyze mass development of the 2008 December 12 CME using the \emph{STEREO} COR1 and COR2 
coronagraphs.We use a well constrained angle of propagation to determine the mass and position of the CME. Combining the mass measurements 
with values for CME velocity and acceleration, the kinetic energy and the magnitude of the force influencing propagation is determined for each point in time.  
Section 2 describes the observations of the event from first appearance of the front in COR1 A and B to the time when the front exits the COR2 A 
and B fields of view. 
Section 3 describes the methods by which the mass, energy, and force are calculated with \emph{a priori} knowledge of the propagation angle. 
Section 4 includes the results and Section 5 discusses the possible forces attributable to the observed accelerations and whether they are 
magnetic or aerodynamic in origin. This is followed by conclusions in Section 6.

\section{Observations}

\begin{figure*}
\includegraphics[scale=0.55, angle=90]{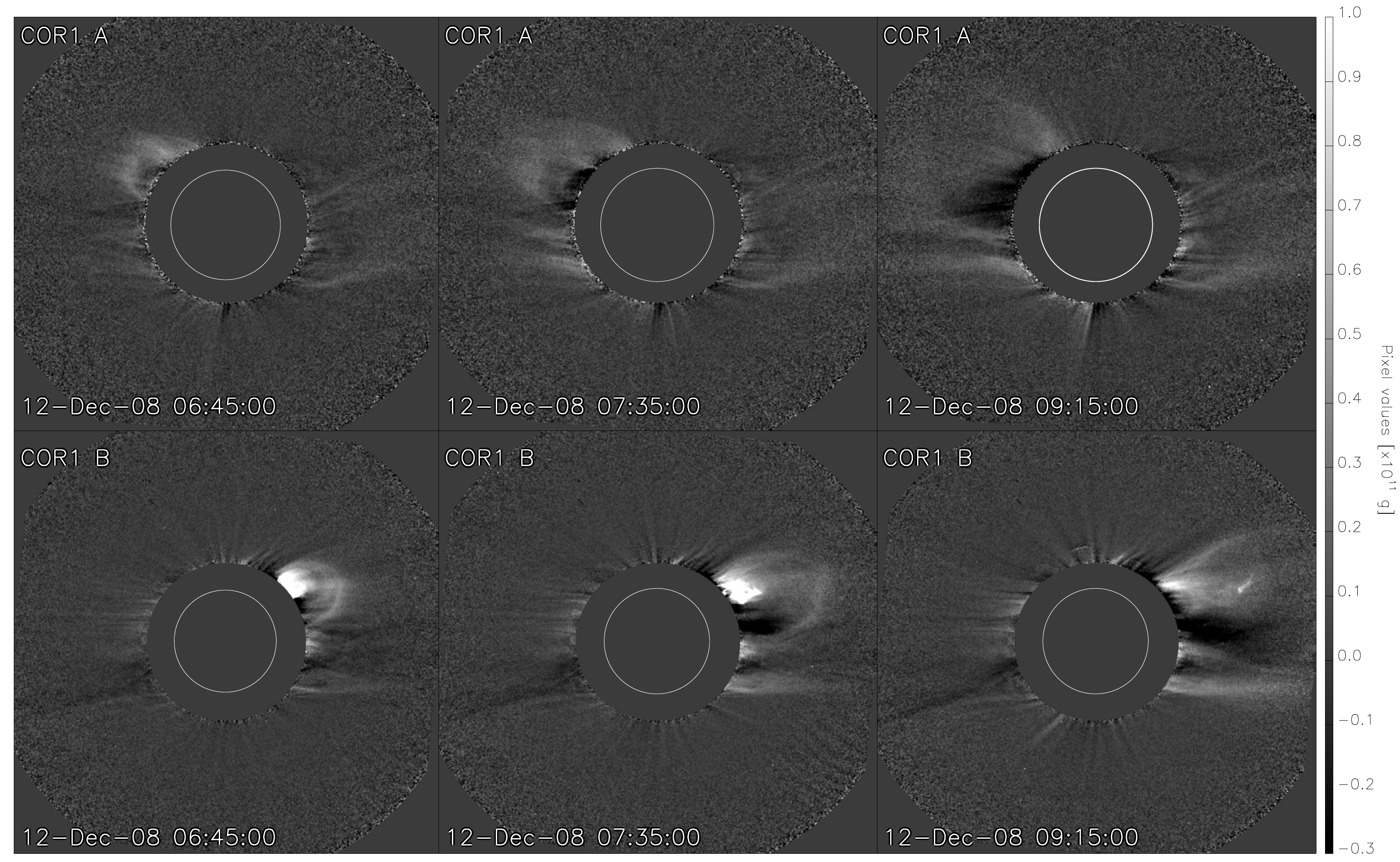}
\caption {Selection of base difference images of the CME in COR1\,A (top row) and COR1\,B (bottom row),  with pixel values of grams. The CME is 
quite faint in the A images and appears not to have as much structure as in B. There is a large contribution to mass from a near-saturated region to 
the upper flank of the CME in the B images. Such saturation in the mass images coincides spatially with the prominence in total brightness images.}
\label{fig:STEREO_COR1A&B}
\end{figure*}

\begin{figure*}
\includegraphics[scale=0.55, angle=90]{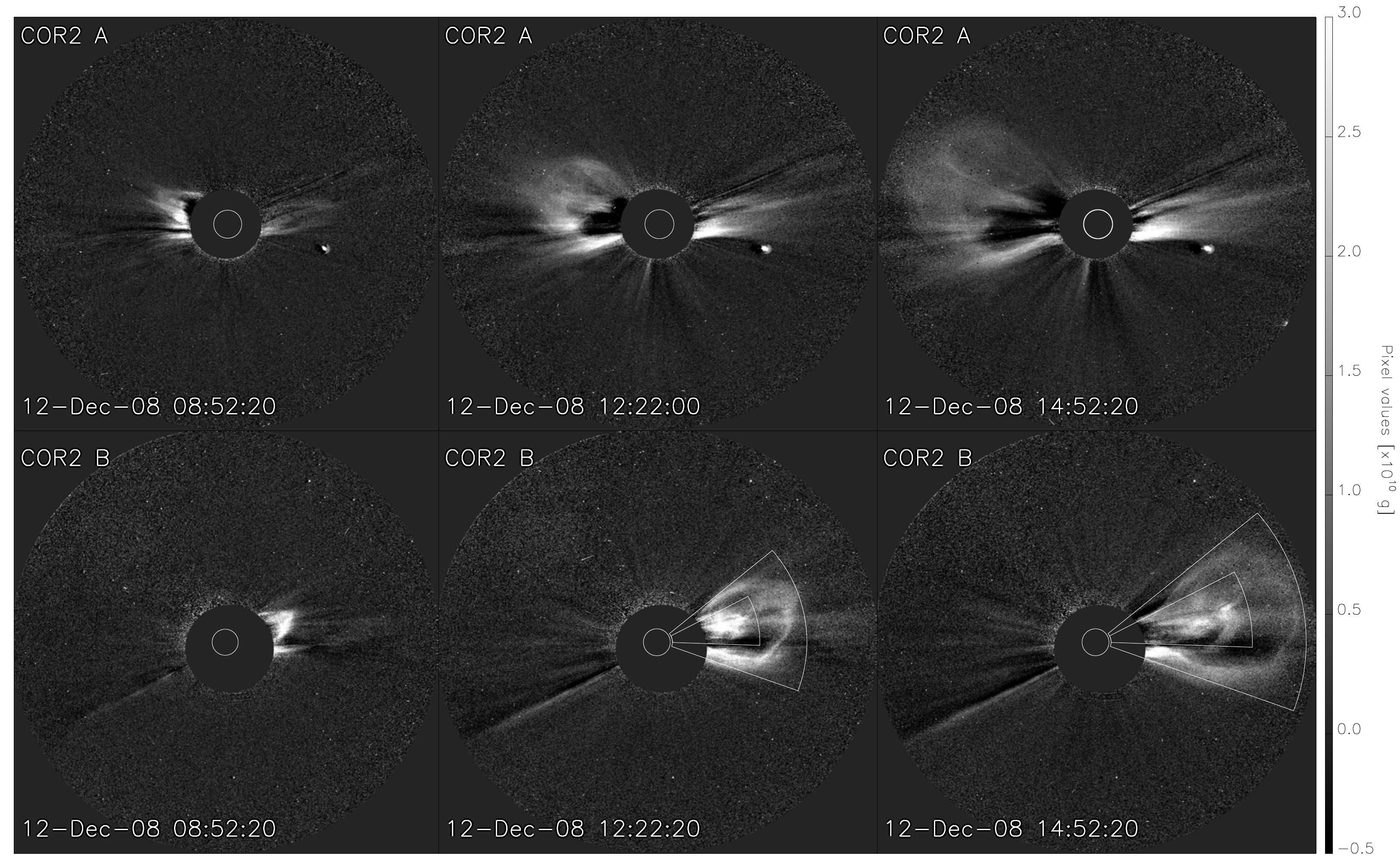}
\caption {Selection of base difference images of the CME in COR2\,A (top row) and COR2\,B (bottom row), with pixel values of grams. The CME is 
clearly distinguishable in both fields of view. Only the B field view shows clearly the three part structure of core, cavity and front. The COR2\,B 
images were used to measure core and front mass separately}
\label{fig:STEREO_COR2A&B}
\end{figure*}

The COR1 images used in this analysis span from 2008 December 12 04:05\,UT to 15:45\,UT, with a cadence of 10\,minutes. The three 
polarization states of COR1 were combined to make total brightness images in units of mean solar brightness (MSB). 
Base difference images were produced using the 04:05\,UT image (in both COR1\,A\,and\,B) as a background to be subtracted from all subsequent 
images. A sample of such images for both COR1\,A and B can be found in Figure~\ref{fig:STEREO_COR1A&B}. The COR2 images analyzed 
range from from 07:22\,UT to 17:52\,UT, with a cadence of 30\,minutes. As with the COR1 images, total brightness images were created for COR2, 
and a set of base difference images were then produced using the 07:22\,UT image as a suitable background. A selected set of images from COR2 
can be found in Figure~\ref{fig:STEREO_COR2A&B}. 
	
	At 04:35\,UT the leading edge of a CME appeared in COR1\,A and B coronagraphs at a height of $\sim$1.4\,$R_{\odot}$, off the east and west 
limb respectively. In COR1\,B the CME first appears as a set of rising loop-like structures followed by a prominence, part of which appears to fall 
back to the surface at 08:00\,UT while the remainder was ejected and follows the rising loop-like structures which eventually become the CME front. 
The rising prominence was not apparent at any stage of the propagation in COR1\,A and the advancing front remains the only distinguishable facet 
of the CME from this line-of-sight (LOS).
		
A noteworthy caveat of using base difference imaging is the assumption that the background corona in the pre-event image has the same 
brightness in all subsequent images. This may not always be true and any excess brightness in the pre-event image will produce negative pixel 
values in the base difference. This is apparent in the COR1 images as the CME interacts with a streamer, displacing it as the leading CME front 
expands laterally as well as moves outward. The streamer is visible as a dark feature that grows with time at the southward flank of the CME in the 
COR1\,B images, Figure~\ref{fig:STEREO_COR1A&B}. The black areas are indicative of negative pixel values. 
The COR1\,A images also suffer from negative pixels, especially at later times, see Figure~\ref{fig:STEREO_COR1A&B} top row, 09:15\,UT image. 
The front of the CME starts to exit both the A and B field of view at $\sim$08:35\,UT.

The CME first appears in the COR2 field of view at $\sim$07:52\,UT with the CME apex at a height of $\sim$3\,$R_{\odot}$ in both A and B 
images. In the B coronagraph, by 10:52\,UT the three part structure of core, cavity, and bright front is clearly visible and the overall structure grows 
in size as the CME propagates to larger heights. The core becomes more tenuous and the mass distribution becomes homogenous after 15:52\,UT 
when the front starts to exit the field of view. The distinction between core and front is not as clear in COR2\,A and the mass distribution appears more 
homogenous throughout the propagation. As with the COR1 images, COR2\,A is also affected by excess brightness in the pre-event image, as is 
apparent by a growing dark feature in its southern half. As the pre-event image for COR2\,B is the cleanest of the pre-event images (it contains the 
least contamination by streamers), the COR2\,B data are considered the best candidate for accurate CME mass measurements.


\section{CME Mass Measurement Methods}
\begin{figure}[h!]
\includegraphics[trim = 0cm 0cm 0cm 0cm, scale=0.3, angle=270]{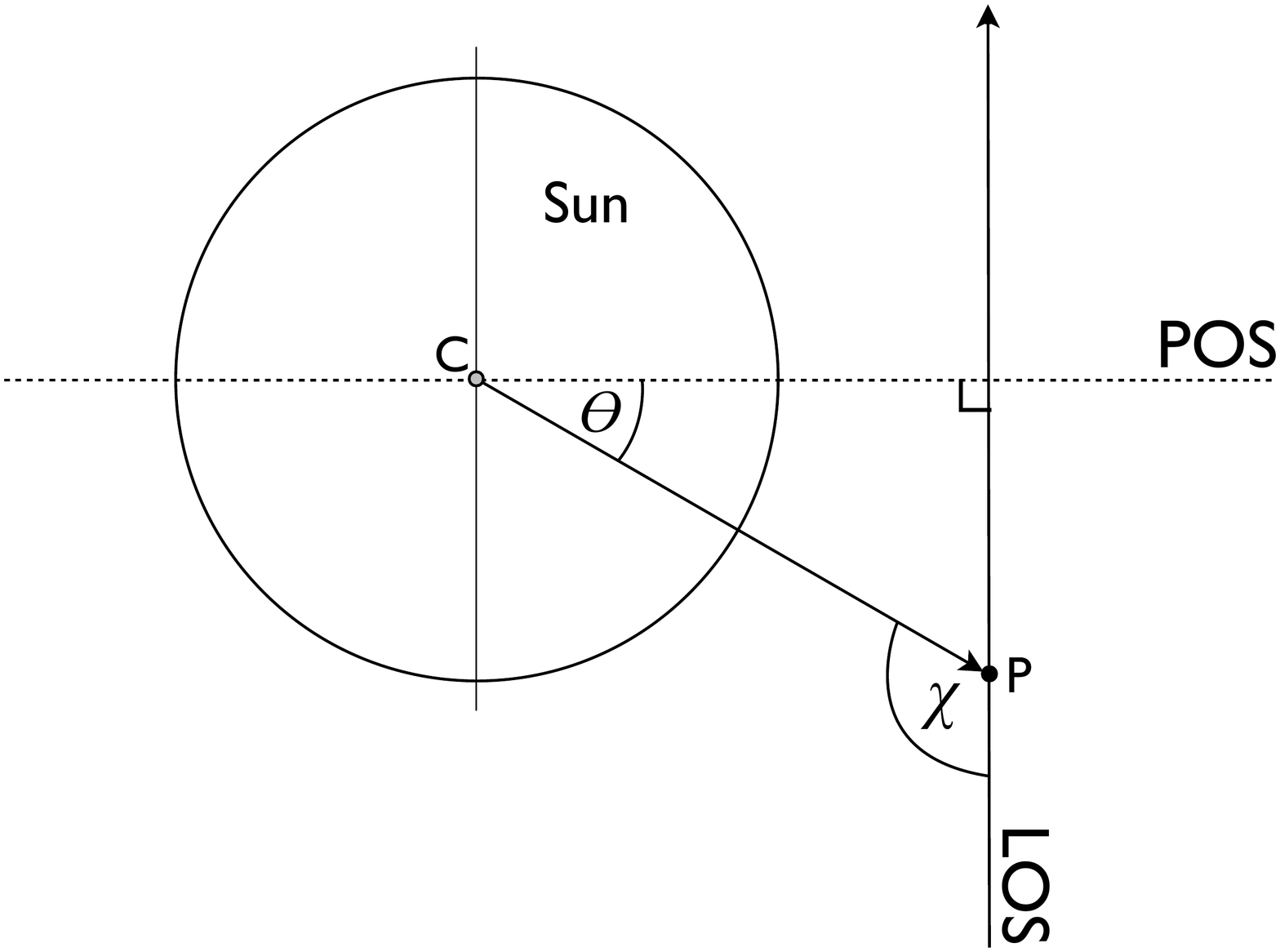}
\caption{Schematic showing the relative orientation of the line-of-sight (LOS), and the plane-of-sky (POS). Electron position is at point P and C is 
Sun center. The vector CP may also represent CME propagation direction. Scattering efficiency is heavily dependent on the angle $\theta$ 
(or $\chi$) and is least efficient 
when $\theta = 0^{\circ}$ ($\chi=90^{\circ})$.}
\label{fig:LOS_POS_2}
\end{figure}

The method by which mass measurements are derived from white light coronagraph images is based on theory first developed by \citet{min30} in 
which the scattering geometry of a single electron at a particular point in the solar atmosphere is considered. Further development of the theory by 
\citet{vdeh50} led to the derivation of what are now known as the van de Hulst coefficients. The coefficients treat each component of the incident 
electric field vector separately and take into account the finite size of the solar disk \citep{min30, 
bil66, how09}. An important fact arising from these expressions is the dependence of scattering intensity on the angle, $\chi$, between the radial 
vector from sun centre to the scattering electron and a position vector from observer towards the electron--the LOS, see Figure~\ref{fig:LOS_POS_2}. 
Scattering efficiency is minimized when this angle is $90^{\circ}$. However, along the LOS such an angle occurs at the point of minimum distance 
from sun centre where the incident intensity (that the electron receives) and electron density are maximized. This means scattered light in the 
corona is most intense along a plane 
perpendicular to the observer's LOS despite the efficiency of scattering being minimized at such viewing angles \citep{how09}. This plane 
perpendicular to the LOS is known as the plane-of-sky (POS)

Studies using single LOS coronagraph data are often hindered by the unknown CME propagation angle from the POS, e.g., unknown $\theta$ (or $\chi$) in 
Figure~\ref{fig:LOS_POS_2}. This leads to the incorrect angle being used when inverting the van de Hulst coefficients to calculate the number of electrons 
contributing to the scattered light. 
Furthermore, because the 3-D extent of the CME is unknown it is also assumed that the CME is confined to the 2-D sky plane, leading to a significant CME 
mass underestimation \citep{vou00}.

The CME of 2008 December 12 was Earth-directed \citep{byr10}, making it roughly the same angular distance from both the \emph{STEREO}\,A
\,and\,B spacecraft, then located $\pm$45 degrees from Earth. This known angle of propagation was used to convert from pixel values of MSB to 
grams via the expression
\begin{equation}
m_{pixel}=\frac{B_{obs}}{B_{e}}\times1.97\times10^{-24}\,\mathrm{g}
\end{equation}
where $B_{obs}$ is the observed MSB of the pixel, $B_{e}$ is the electron brightness calculated from the van de Hulst coefficients, and 1.97$
\times10^{-24}$\,g is a factor that converts the number of electrons to mass, assuming a completely ionized corona with a composition of 90\% 
hydrogen and 10\% helium. The known angle of propagation allowed the correct value of $B_{e}$ to be computed resulting in a significant 
reduction in the uncertainties associated with the propagation angle. The largest remaining uncertainty is the unknown angular width along 
the LOS. This uncertainty was quantified in a similar approach to the method outlined in \citet{vou00}. This simulates the brightness of a CME with 
homogeneous density distribution and finite angular width along the LOS--longitudinal angular width $\Delta$$\theta_{long}$, allowing calculation 
of a simulated observed mass. Comparing this to the actual mass allowed for an evaluation of CME mass underestimation for given values of $
\Delta$$\theta_{long}$.  Since the values for $\Delta$$\theta_{long}$ are unknown, the expression derived in \citet{byr10} for the \emph{latitudinal} 
angular width of this CME as a function of height, $\Delta$$\theta_{lat}$$(r)=25r^{0.22}$, was used to define an upper limit to $\Delta$$\theta_{long}
$. It was assumed the CME longitudinal angular width is no more than twice the latitudinal angular width, or $\Delta$$\theta_{long}$$\leqslant$\,2$
\times$$\Delta$$\theta_{lat}$. Such an upper limit is in agreement with simulations of flux-rope CMEs which give a typical aspect ratio of broadside 
to axial angular extents of 1.6\,--\,1.9 \citep{krall2006}. Hence the value for $\Delta$$\theta_{long}$ at each height was used to obtain the simulated 
mass underestimation estimates described above. The heights and angular widths used in this study produced CME mass underestimation 
estimates of between 5--10\% for finite angular width uncertainty. An extra mass uncertainty of 6\% was added to account for the assumption of 
coronal abundance of 90\% hydrogen and 10\% helium which can lead to slight errors while converting from pixel values of MSB to grams \citep
{vour2010}. 

To calculate the CME mass a user-selected area (the extent of the CME, for example) of the base difference image was chosen and the pixel 
values within this area were summed to obtain total mass. Figure~\ref{fig:STEREO_COR2A&B} COR2 B images show an example of the sector over 
which pixels were summed (the smaller sectors indicate a different summing region used at a later stage).  
The selected area was chosen for each image in the time sequence of CME propagation so as to determine the mass variation with height in COR1 
and 2 using both A and B. The selection of an area by a point and click method is of course a subjective identification of the the extent of the CME, 
so it is susceptible to user-generated uncertainties. To quantify these uncertainties the mass was obtained for each coronagraph image in the time 
sequence (as described above) and the process was repeated five times in order to obtain the mean CME mass for each image and the standard 
error on the mean. This standard error was defined as the uncertainty due to user bias in the point and click method of CME identification.
The height at each measurement interval was taken to be the heliocentric distance of the CME apex in the image i.e., the apex of the front was chosen 
by simple point-and-click method. The uncertainty on the apex height was also found by the standard error on five runs.

The deflection of a small streamer during CME propagation produces negative pixels in the base difference images. The effect is particularly 
apparent in the COR1 images, Figure~\ref{fig:STEREO_COR1A&B}. It is difficult to unambiguously distinguish between streamer and CME, making 
it difficult to quantify the uncertainty introduced due to streamer interaction. To make an estimate of the streamer's effects, a calculation of its mass 
in the pre-event image was made. A number of different samples of the area of the streamer in the COR1\,B pre-event image that effects all 
subsequent images produced a mass estimate of $\sim$5$\times$10$^{14}$\,g. This mass was used as a measure of the uncertainty introduced 
due to streamer interaction in the COR1\,B images. A similar analysis of the COR1\,A pre-event images gave a streamer mass estimate of 
$\sim$7$\times$10$^{14}$\,g. COR2 images are relatively unaffected by significant changes in background coronal brightness and do not suffer 
from negative pixel values to as large an extent as COR1. The pre-event image of COR2\,B is particularly clean and free of background streamers, 
hence COR2\,B images are considered to provide most accurate CME mass estimation.

Finally, in order to obtain a more complete and continuous estimate of CME mass growth, the masses determined from both COR1 and COR2 
coronagraphs were summed in those cases where image times of the inner and outer coronagraphs overlapped\footnote{A difference in cadence of 
the inner and outer coronagraphs means that the images closest in time have a three minute offset e.g., a COR1 image taken at 07:25 UT was 
considered to be coincident in time with the COR2 image at 07:22 UT}. The overlap in the inner and outer corongraphs' fields of view was also 
taken into account in this summation.

A concise measurement of the CME kinematics, such as velocity and acceleration, were taken from the results of the study of \citet{byr10}. Since 
these kinematics take into account the true three dimensional surface of the front they provide reliable estimates of CME velocity and acceleration 
in 3-D space. These velocity and acceleration measurements were used in the calculation of kinetic energy and total force on the CME for each 
point in time. The CME mass used in all energy and force calculations was the asymptotic mass it approaches at later stages of its evolution 
beyond 10\,$R_{\odot}$ as observed from the \emph{STEREO B} spacecraft i.e.,\,3.4\,$\pm$\,1.0$\times$10$^{15}$\,g. As will be shown, there is 
good motivation for the use of constant mass in the magnitude of kinetic energy and force estimates.


\section{Results} \label{bozomath}

\subsection{CME Mass Estimates}
\begin{figure}
\includegraphics[scale=0.55, angle=0]{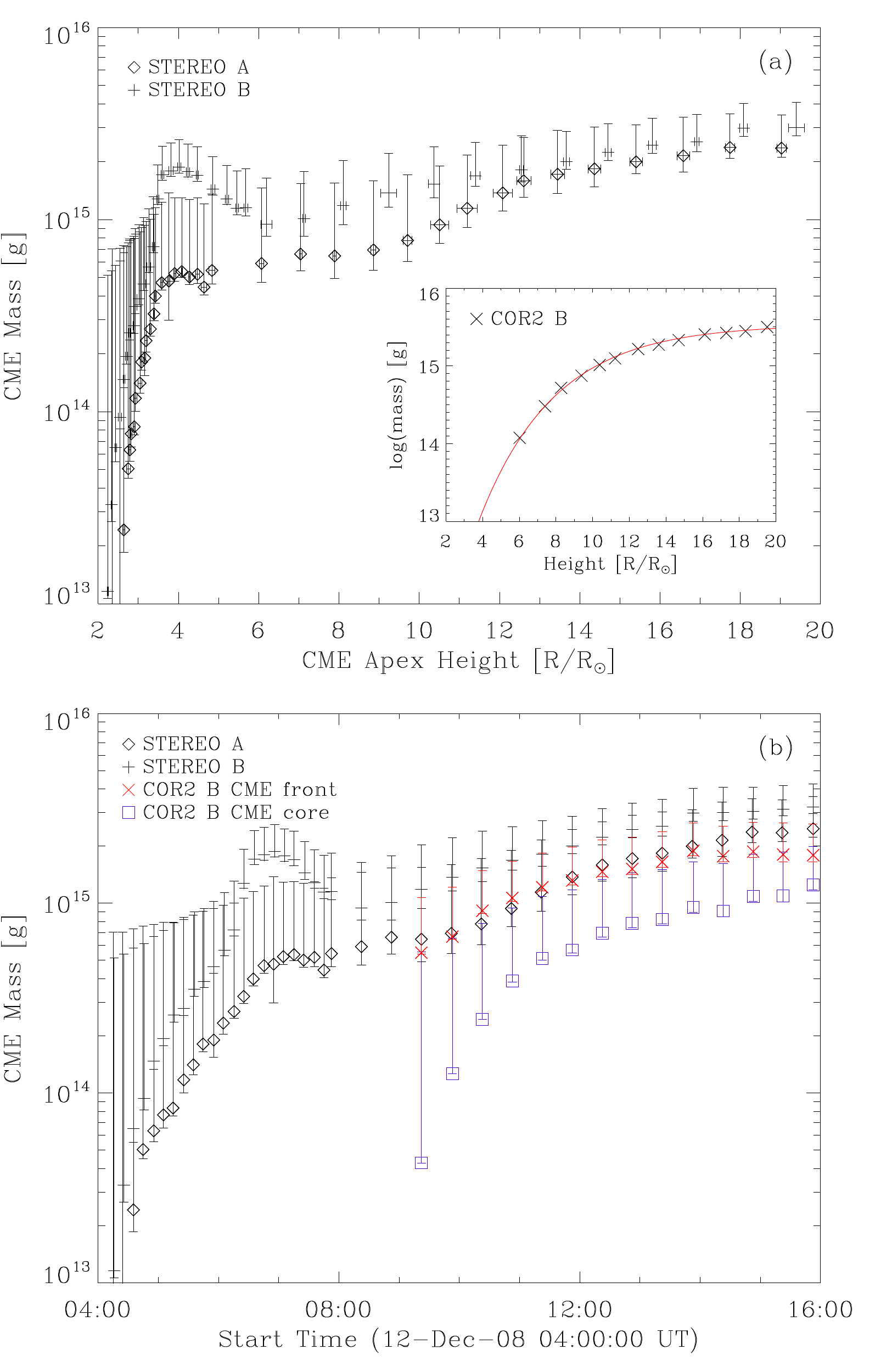}
\caption{CME mass development with height (a) and time (b), for the 2008 December 12 CME. After $\sim$08:00 UT ($\gtrsim$5\,$R_{\odot}$) the 
masses from the inner and outer coronagraphs are summed to show uninterrupted 
mass development from $\sim$2--20\,$R_{\odot}$ over a period of 12 hours. The small bump in the 
CME mass at $\sim$07:00\,UT ($\sim$4\,$R_{\odot}$) is probably due to an unknown amount of H-$\alpha$ emission from the prominence. Mass of 
CME front and core are also shown, red $\textquoteleft$$\times$' and blue square, for COR2\,B, panel (b). After 14:52\,UT they share approximately 
equal mass. The inset of (a) shows mass development with height for COR2\,B only; the red curve represents a fit to the data whereby the mass 
asymptotically approaches $3.4\,\pm\,1.0\times10^{15}$\,g. }
\label{fig:20081212_mass_ht}
\end{figure}

The results of the calculation for CME mass development with time and height for both \emph{STEREO} A and B coronagraphs are shown in 
Figure~\ref{fig:20081212_mass_ht}. In panel (a), the height values are those taken from a point-and-click method of tracking the CME apex; these 
heights are corrected for CME propagation angle of $\sim$$45^{\circ}$.  In both panels (a) and (b), the mass estimates of \emph{STEREO} A and 
B follow a similar trend and have similar values at each stage in the propagation. Such good agreement between mass values is a good indicator 
that $\sim$$45^{\circ}$ is the correct angle of propagation from the sky-plane. A change in the cadence of mass measurements is noticeable at 
$\sim$08:00 UT (or $\gtrsim$5\,$R_{\odot}$). This is due to the use of only COR1 images (with a cadence of 10\,minutes) prior to this time, and 
the use of the COR2 plus COR1 images after this time (the cadence of these measurements follows that of COR2\,--\,30\,minutes). Comparing 
A and B below 4.5\,$R_{\odot}$, mass values show a similar trend and increase at the same rate, but at approximately 3\,$R_{\odot}$ the mass 
measurements in COR1\,B appear to increase to a much larger value then fall again. This effect is visible in the COR1\,A measurements, albeit 
diminished. It is probably due to the presence of a prominence which contains a significant mass content and therefore contributes a large amount
to total measured CME mass. Also, early on in its propagation, the prominence may still be emitting H-$\alpha$ line radiation (656.28 nm) due 
to the larger fraction of neutral hydrogen at its cooler temperatures.
The COR1 imaging passband is centered on H-$\alpha$ so any emission in the prominence from neutral hydrogen could be contributing to light 
received by the COR1 coronagraphs, this is apparent from the saturation region in the COR1\,B images in Figure~\ref{fig:STEREO_COR1A&B}. Since this is resonance line 
emission, and not Thomson scattered emission, it leads to an erroneous measurement in CME mass. Thus, it is assumed the larger rise and fall in 
CME mass is caused by the prominence entering and exiting the COR1\,B field of view. The effect is diminished in COR1\,A since the prominence 
does not enter the FOV to as large an extent as in COR1\,B. The interpretation that the $\textquoteleft$mass bump' is not actual mass growth (or 
loss) is supported by previous measurements where CME mass increase follows a trend with height described by $M_{cme}(h)=M_{a}(1-e^{-h/h_a})
$, where $M_a$ is the final mass the CME approaches asymptotically and $h_a$ is the height at which the CME reaches 0.63$M_a$ \citep
{cola09}, with no  $\textquoteleft$bump'~in mass earlier on. The decline in mass after the peak may be explained by the ionization of neutral 
hydrogen such that H-$\alpha$ emission diminishes and simply becomes Thomson scattering of free electrons, as with the rest of the CME 
material. 

In order to produce a fit to the data, the COR2\,B mass results were chosen because its pre-event image was largely free of any bright streamers or 
other features which introduce unwanted effects in the production of base difference images, as described above. A fit with the above equation 
resulted in a final asymptotic CME mass of 3.4\,$\pm$\,1.0$\times$10$^{15}$\,g, with a scale height of $h_a=2.9\,R_{\odot}$. This fit is plotted 
along with the COR2\,B data in the inset panel of Figure~\ref{fig:20081212_mass_ht}(a). Note that the mass increase is due to material coming 
up from below the occulting disk, and not actual mass gain of the CME. The uncertainty on the above asymptotic mass value was taken to be 30\%, from the 
largest uncertainty  due to finite width, the conversion factor uncertainty as described above, the standard error user-generated uncertainty, and uncertainty due 
to streamer interaction.

In each image where the CME core and front are distinguishable, their masses were measured separately. This was carried out by user selected 
regions demarcating the areas of core and front, see COR2\,B at 12:22\,UT and 14:52\,UT in Figure~\ref{fig:STEREO_COR2A&B} for an example of the separate core and front 
sectors over which pixel values were summed to obtain total mass. The uncertainties due to finite width of the observed object also apply to the 
core and front measurements, however, since the widths of these particular areas of the CME are unknown we chose the maximum uncertainty of 
10\% from the above analysis since neither core nor front can be any wider than the maximum width assigned to this CME. The remaining 
uncertainties described above were also applied. The mass development of core and front with time is shown in Figure~\ref{fig:20081212_mass_ht}(b). The two mass 
measurements are subject to an observational effect of apparent exponential mass growth, however by the time the CME is fully in the field of view 
at 14:45\,UT the core and front share approximately equal mass. 

\subsection{CME Forces and Energetics}

In the following calculations, all measurements of force and kinetic energy use the asymptotic mass of 3.4\,$\pm$\,1.0$\times$10$^{15}$\,g and 
not the instantaneous mass values calculated from each coronagraph image i.e., the CME is considered to begin its propagation with this mass and 
does not acquire any mass as it propagates. 

\begin{figure}[h!]
\includegraphics[scale=0.75, angle=0]{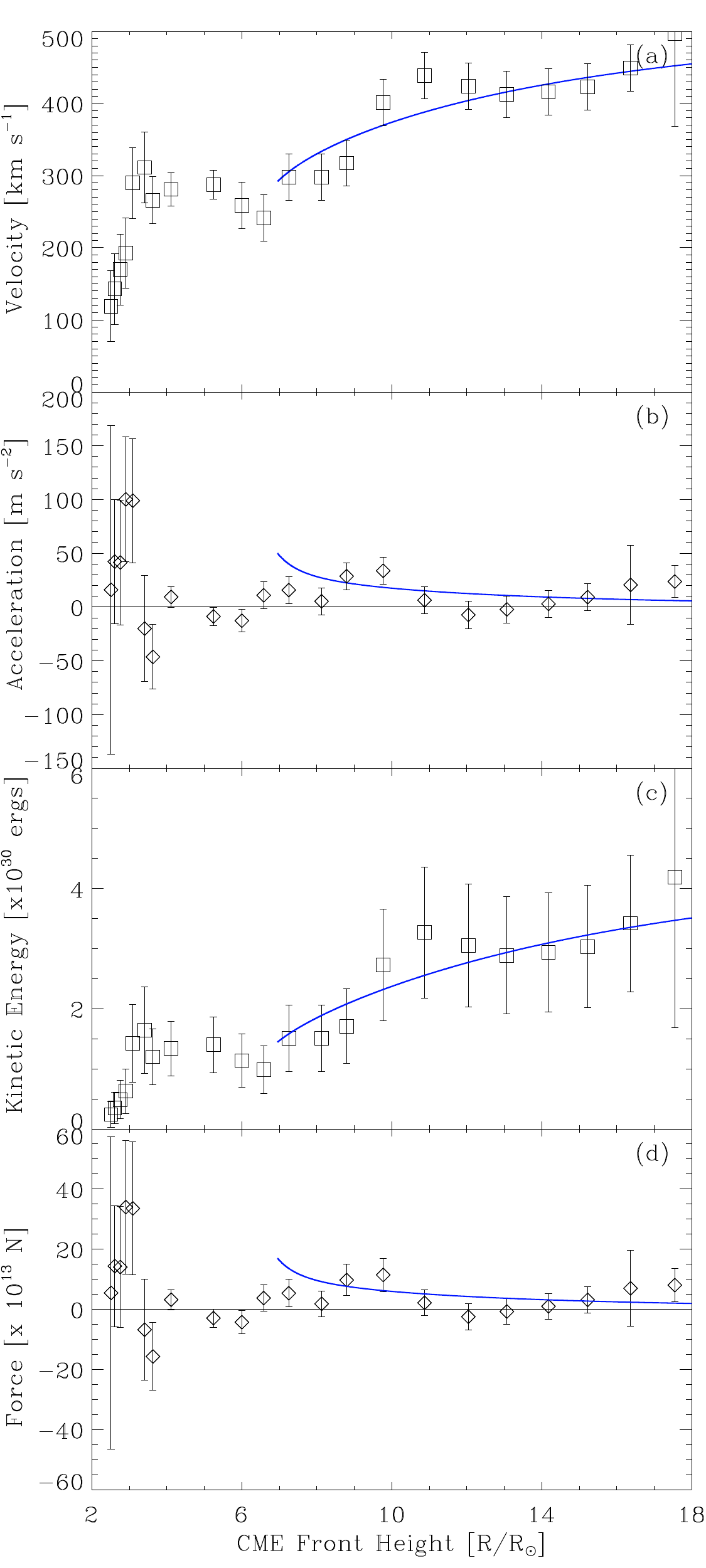}
\caption{(a) CME velocity as a function of heliocentric distance, including a fit to the data produced using an aerodynamic drag model beyond $\sim$7\,$R_{\odot}$ \citep{byr10}. (b) Acceleration of CME, including fit,  derived from the velocity data and fit. 
Panel (c) and (d) show the kinetic energy and force, respectively, both calculated using constant CME mass of $3.4\,\pm\,1.0\times10^{15}$\,g and kinematics results from (a) and (b). Also shown are the fits to energy and force produced from fits to velocity and acceleration.}
\label{fig:force_20081212}
\end{figure}
Estimates of the force and kinetic energy use the 3-D velocity and acceleration measurements produced by \citet{byr10}. Their method firstly identifies the CME front in each coronagraph image using a multiscale edge detection filter. The front edges were then used to define a quadrilateral in space into which an ellipse is fit, this method is known as elliptical tie-pointing. This was done for multiple horizontal planes through the CME so that the fit ellipses outline a curved front in 3-D space.
The speed and acceleration were then deduced from the change in position of the front, with time, through the \emph{STEREO} COR1, 
COR2 and HI fields of view. Since mass measurements in this study use only the COR1 and COR2 coronagraphs, HI kinematics measurements have been excluded here. The CME front position uncertainty in \emph{STEREO A}\,and\,\emph{B} coronagraphs was determined from the filter width in the multiscale analysis. Velocity and acceleration uncertainties were then propagated from position uncertainty.  Figure~\ref{fig:force_20081212}(a) shows CME velocity as a function of heliocentric distance, along with acceleration in panel (b). 

The CME kinetic energy was calculated using $E_{kin}=1/2M_{cme}v_{cme}^{2}$, where $M_{cme}$ is the final asymptotic mass of 3.4\,$\pm$\,1.0$\times
$10$^{15}$\,g and $v_{cme}$ are the instantaneous velocity measurements, results of this calculation are shown in Figure~\ref{fig:force_20081212}(c). The kinetic energy shows 
an initial rise towards 6.3\,$\pm$\,3.7$\times$10$^{29}$\,ergs at $\sim$3\,$R_{\odot}$, beyond which it rises steadily to 4.2\,$\pm$\,2.5$\times
$10$^{30}$\,ergs at $\sim$18\,$R_{\odot}$, these values are similar to those reported in \citet{vou00,vour2010} and \citet{emslie2004}. 

The total force on the CME was calculated using $F_{total}=M_{cme}a_{cme}$, where $M_{cme}$ is as above and $a_{cme}$ is taken from the 
instantaneous acceleration values.  
As shown in panel (d) of Figure~\ref{fig:force_20081212}, the force initially grows significantly, reaching a maximum value of 3.4\,$\pm$\,2.2$\times$10$^{14}$\,N at $\sim$3\,$R_
{\odot}$. The early rise and fall in acceleration (or force) is in agreement with a previous study of a CME observed to reach peak acceleration at $\sim$1.7\,$R_{\odot}$ after which it reaches a constant velocity beyond $\sim$3.4\,$R_{\odot}$ \citep{gallagher03}.  Such results are also found in a statistical study 
which shows that the majority of CMEs have peak acceleration in the low corona with a mean height of maximum acceleration at 1.5\,$R_{\odot}$ 
\citep{bein2011}. Similarly, observational studies by \citet{zhang2001} and \citet{zhang2004} also show early phase peak acceleration between 
2--5\,$R_{\odot}$ and forces on the order of 10$^{15}$\,N and 10$^{12}$\,N, depending on whether the CME shows large initial acceleration or a 
slow, more gradual acceleration.

After this early peak, the force drops to an average value of 3.8$\pm$5.4$\times$10$^{13}$\,N at distances between 7--18\,$R_{\odot}$.
It is apparent from Figure~\ref{fig:force_20081212}(a) that the velocity continues to increase beyond $7\,R_{\odot}$, implying that a positive radial force must be present. To clarify this, a fit to the velocity data using a model for solar wind drag on the CME beyond $7\,R_{\odot}$ (as outlined in \citet{byr10}) is shown in Figure~\ref{fig:force_20081212}(a). Although the data suggest a non-monotonic increase in velocity, the fit reveals that propagation is best described by a steadily increasing velocity between 7--18\,$R_{\odot}$. The acceleration and kinetic energy curves derived from this velocity fit are shown in Figure~\ref{fig:force_20081212}(b) and (c). In Figure~\ref{fig:force_20081212}(d), the curve for the force derived from the velocity fit initially deviates from the data at $\sim$7\,$R_{\odot}$, however beyond this distance there is good agreement with the data and the derived force is entirely positive.  This suggests that the solar wind exerts a positive aerodynamic drag force on the CME, resulting in a velocity that approaches the asymptotic solar wind speed at large heliospheric distances.

\section{Discussion} \label{bozomath}

It should be noted that Figure~\ref{fig:20081212_mass_ht} shows an overall exponential increase in CME mass with height which could be interpreted as the CME rapidly 
gaining mass as it propagates. Care should be taken with this interpretation since this apparent exponential mass increase is almost certainly due 
to the CME moving into the field of view, therefore allowing us to measure more of its mass content; such an interpretation is in agreement with 
similar assertions made in \citet{vour2010}. 
It is difficult to distinguish between actual CME mass growth and an apparent growth due to more of the CME being observed. If the initial early rise
in CME mass is assumed to be an observational artifact then we can interpret the CME mass to be in the range of (3--3.5)$\times$10$^{15}$\,g for 
most of its early
propagation i.e., the CME already has such a mass before launch and does not acquire more mass (via inflows or otherwise) during propagation.
Such an interpretation is in agreement with CME mass measurements calculated from dimmings in \emph{STEREO} Extreme Ultraviolet Image
(EUVI) images, which show the mass calculated from EUV images to be approximately equal to CME mass in COR2 images,  $m_{EUVI}/m_{COR2} =1.1\pm0.3$  \citep{aschw09}. Once the CME bubble is in the field of view at $\sim$10\,$R_{\odot}$ the mass in its entirety can be measured
and the increase beyond this point, if any, is slow and steady, Figure~\ref{fig:20081212_mass_ht}.

The early stages of CME propagation are dominated by a sharp rise to a peak force of 3.4\,$\pm$\,2.2$\times$10$^{14}$\,N at $\sim$3\,$R_
{\odot}$ followed by a sharp decline, Figure~\ref{fig:force_20081212}(d). The catastrophe model \citep{forbes1991,forbes1995,lin2000}, 
magnetic breakout model \citep{antio99,lynch2008}, and toroidal instability model \citep{chen1996,kleim2006} employ a number of forces acting 
on the CME to produce an over all acceleration into interplanetary space. For example, the toroidal instability model used by \citet{chen1996} uses 
a Lorentz hoop force (or Lorentz self-force), solar wind drag, and gravity to provide a net force acting on the CME between 2--3\,$R_{\odot}$ that quickly 
rises to a peak total force of $\sim$10$^{16}$\,N and then falls rapidly.

If we assume that the peak force observed for the 2008 December 12 CME is the net force due to similar forces used in the above models, such as 
the solar wind drag, gravity, and some form of magnetic CME driver e.g., a $\vec{J}\,\times\,\vec{B}$ force, we may estimate their relative 
contribution.
The force due to solar wind drag on the CME is given by
\begin{equation}
\vec{F}_d=-\frac{1}{2}C_{d}\rho_{sw}A_{cme}(\vec{v}-\vec{v}_{sw})\mid\,\vec{v}-\vec{v}_{sw}\mid
\end{equation}
where $M_{cme}$ is the CME mass, $\vec{v}$ is the CME velocity, $C_{d}$ is the drag coefficient, $\rho_{sw}$ is the solar wind mass density, $A_
{cme}$ is the CME area exposed to solar wind drag and $\vec{v}_{sw}$ is the solar wind velocity \citep{malo10}. To estimate the effects of this force 
we use $\rho_{sw}=n_{p}m_{p}$, where $m_{p}$ is proton mass, and assume ionization fraction of $\chi$\,=\,1 such that $n_{p}=n_{e}$\,$[cm^{-3}]$. 
Electron density, and hence proton density, is then given by an interplanetary density model derived from a special solution of the Parker solar wind 
equation \citep{Mann1999}, solar wind velocity values as a function of height are also determined using this model. $A_{cme}$ is estimated using 
the expression derived in \citet{byr10} for latitudinal angular width of the CME as a function of height, $\Delta$$\theta_{lat}$$(r)=26r^{0.22}$. This is 
used to derive an arc length of the CME front and, as above, making the assumption $\Delta$$\theta_{long}$$=$\,2$\times$$\Delta$$\theta_{lat}$,
the two arc lengths derived from these angles then give the surface that the solar wind acts on, thus $A_{cme}$\,=\,1352$r^{2.44}$. Setting the drag 
coefficient $C_{d}=1$, and using the \citet{Mann1999} model to derive a density and a solar wind velocity of 2.3$\times$10$^{5}$\,cm$^{-3}$  and 
70\,km\,s$^{-1}$, respectively, equation [1] then gives a force of $\vec{F}_{d}=-8.0\times10^{12}\,\hat{r}\,$\,N for solar wind drag at $\sim$3\,$R_
{\odot}$, where $\hat{r}$ is a unit vector in the positive radial direction. 

A simple estimate of force due to gravity is given by $\vec{F}_{g}=GM_{\odot}M_{cme}/\vec{r}\,^2$, where $G$ is the universal gravitational 
constant, $M_{\odot}$ is solar mass, $M_{cme}$ is CME mass, and $\vec{r}$ is a heliocentric position vector\footnote{Ideally the heliocentric 
distance of the CME centre of mass would be used here. However an unknown amount of mass is obscured by the coronagraphs occulting disk, 
making the mass distribution and hence COM difficult to determine. Thus the CME front height is used in the calculation of force due to gravity}. 
Given a CME mass of 3.4$\times$10$^{15}$\,g the force due to gravity at a heliocentric distance of 3\,$R_{\odot}$ is $\vec{F}_{g}=-1.0\times10^
{14}\,\hat{r}$\,N. The only remaining contribution is due to some form of magnetic CME driver, $F_{mag}$, which is estimated using 
\begin{equation}
\vec{F}_{mag}= \vec{F}_{total}-\vec{F}_{d}-\vec{F}_{g}
\end{equation}
(the pressure gradient in the CME equation of motion is assumed to be negligable and has been omitted here). Using the above values, the total 
magnetic contribution to CME force is calculated to be $\vec{F}_{mag}\approx4.5\times10^{14}\,\hat{r}$\,N at 3\,$R_{\odot}$, indicating that this is 
the largest driver of CMEs at low coronal heights. Lorentz force dominated dynamics in early phase CME propagation are reported in \citet
{bein2011}, in which a statistical study of a large sample of CMEs in EUVI, COR1, and COR2 indicated an early phase acceleration for the majority 
of CMEs that is attributable to a Lorentz force.  A similar result of an observational study by \citet{vrs06} found that the Lorentz force plays a 
dominant role within a few solar radii. It should be noted that although we have labelled the force $F_{mag}$, there is no distinction on the exact 
form of this force e.g., whether it is magnetic pressure, magnetic tension, or a Lorentz self-force that acts as the driver. Also, any non-radial motion 
of the CME, such as that described in \citet{byr10}, is not taken into account here; any force estimates are purely 
radial in direction.

 \section{Conclusion} \label{bozomath}
 The \emph{STEREO} COR1/2 coronagraphs have been used to determine the mass development of the 2008 December 12 CME. Knowledge of 
the longitudinal propagation angle of the CME allowed for a significant reduction in the mass uncertainty, giving a final estimate of 
3.4\,$\pm$\,1.0$\times$10$^{15}$\,g. Using kinematics results of a previous study \citep{byr10}, the velocity and acceleration of the CME were
 combined with the mass measurements to determine the kinetic energy and total force on the CME. The early phase propagation of the CME was found to be 
dominated by a force of peak magnitude of 3.4\,$\pm$\,2.2$\times$10$^{14}$\,N at $\sim$3.0\,$R_{\odot}$, after which the magnitude declines 
rapidly and settles to and average of 3.8 $\pm$ 5.4$\times$10$^{13}$\,N. This early rise and fall in total force (or acceleration) is in agreement 
with previous observations of CME kinematics \citep{gallagher03, bein2011}. Similarly results of observational studies by \citet{zhang2001} and 
\citet{zhang2004} also show early phase peak acceleration between 2--5\,$R_{\odot}$ and forces on the order of 10$^{15}$\,N and 10$^{12}$
\,N. The kinetic energy shows an initial rise towards 6.3\,$\pm$\,3.7$\times$10$^{29}$\,ergs at $\sim$3\,$R_{\odot}$, beyond which it rises 
steadily to 4.2\,$\pm$\,2.5$\times$10$^{30}$\,ergs at $\sim$18\,$R_{\odot}$, such order of magnitudes are similar to those reported in \citet
{vou00,emslie2004} and are typical of CME kinetic energies \citep{vour2010}.

Such CME kinematics and dynamics property estimates cannot be carried out when unknown propagation angle hinders an accurate calculation of 
CME mass, hence adding unacceptable uncertainty to any subsequent calculations. This highlights the need for similar studies using the \emph
{STEREO} mission's ability to accurately determine the physical properties of CMEs, such as mass, with remarkably reduced uncertainty. 
Increasing the accuracy of force estimates of other well studied CMEs will allow for a more complete view of the magnitude of the forces influencing 
CME propagation and will allow model parameters to be more accurately constrained.

\vspace{5mm} 

This work is supported by the Irish Research Council for Science, Engineering and Technology (IRCSET). We also extend thanks and appreciation 
to the \emph{STEREO}/SECCHI consortium for providing open access to their data.
 
\bibliographystyle{apj.bst}
\bibliography{ecarley_review_apj_arxiv.bib}

\end{document}